\documentstyle[a4,12pt]{article}
\renewcommand{\baselinestretch}{1.17}

\begin{document}
\parskip=5pt plus 1pt minus 1pt

\vspace{0.2cm}

\begin{flushright}
{\bf CERN-TH/97-201} \\
{\bf DPNU-97-39}
\end{flushright}

\vspace{0.2cm}

\begin{center}
{\Large\bf On the Parametrization of Flavor Mixing \\ in the Standard Model}
\end{center}

\vspace{0.3cm}

\begin{center}
{\bf Harald Fritzsch}  \footnote{Electronic address:
bm@hep.physik.uni-muenchen.de} \\
{\it Theory Division, CERN, CH--1211 Geneva 23, Switzerland; and} \\
{\it Sektion Physik, Universit$\ddot{a}$t M$\ddot{u}$nchen,
80333 M$\ddot{u}$nchen, Germany}
\end{center}

\begin{center}
{\bf Zhi-zhong Xing} \footnote{Electronic address: xing@eken.phys.nagoya-u.ac.jp} \\
{\it Department of Physics, Nagoya University, Nagoya 464-01, Japan}
\end{center}

\vspace{1.5cm}

\begin{abstract}
It is shown that there exist nine different ways to describe the
flavor mixing, in terms of three rotation angles and one
$CP$-violating phase, within the standard electroweak theory of six
quarks. For the assignment of the complex phase 
there essentially exists a continuum of possibilities, if one allows
the phase to appear in more than four elements of the mixing
matrix. If the phase is restricted to four elements, the phase
assignment is uniquely defined. If one imposes the constraint that the 
phase disappears in a natural way in the chiral limit in which the
masses of the $u$ and $d$ quarks are turned off, only three of the
nine parametrizations are acceptable. In particular the
``standard'' parametrization advocated by the Particle Data Group is 
not permitted. One parametrization, in which the $CP$-violating phase
is restricted to the light quark sector, stands up as the most
favorable description of the flavor mixing.
\end{abstract}

\vspace{0.5cm}
\begin{center}
PACS number(s): 12.15.Hh, 11.30.Er, 12.15.Ff
\end{center}

\newpage

In the standard electroweak theory, the phenomenon of flavor
mixing of the quarks is described by a $3\times 3$ unitary matrix, the
Cabibbo-Kobayashi-Maskawa (CKM) matrix \cite{Cabibbo63,KM73}. This
matrix can be expressed in terms of four parameters, which are usually 
taken as three rotation angles and one phase. A number of
different parametrizations have been proposed in the literature
\cite{KM73}--\cite{FX97}. Of course, adopting a particular parametrization
of flavor mixing is arbitrary and not directly a physical
issue. Nevertheless it is quite likely that the actual values of 
flavor mixing parameters (including the strength of $CP$ violation),
once they are known with high precision, will give interesting information 
about the physics beyond the standard model. Probably at this point it 
will turn out that a particular description of the CKM matrix is more
useful and transparent than the others. For this reason, we find it
useful to analyze all possible parametrizations and to point out their
respective advantages and disadvantages. This is the main purpose of this short note.

In the standard model the quark flavor mixing arises once the up- and
down-type mass matrices are diagonalized. The generation of quark masses 
is intimately related to the phenomenon of flavor mixing. In
particular, the flavor mixing parameters do depend on the
elements of quark mass matrices. A particular structure of the
underlying mass matrices calls for a particular choice of the
parametrization of the flavor mixing matrix. For example, in
Ref. \cite{Fritzsch79} it was noticed that a rather special form of
the flavor mixing matrix results, if one starts from Hermitian mass
matrices in which the (1,3) and (3,1) elements vanish. This has been
subsequently observed again in a number of papers
\cite{Hall}. Recently we have studied the exact form of such a
description from a general point of view and pointed out some
advantages of this type of representation in the discussion of flavor
mixing and $CP$-violating phenomena \cite{FX97}. One 
of the aims of this work is also to view this parametrization in the
context with other ways of describing the flavor mixing.

In the standard model the weak charged currents are given by 
\begin{equation}
\overline{(u, ~ c, ~ t)}^{~}_L \left ( \matrix{
V_{ud} 	& V_{us}	& V_{ub} \cr
V_{cd}	& V_{cs}	& V_{cb} \cr 
V_{td}	& V_{ts}	& V_{tb} \cr} \right ) 
\left ( \matrix{
d \cr s \cr b \cr} \right  )_L \; ,
\end{equation}
where $u$, $c$, ..., $b$ are the quark mass eigenstates, $L$ denotes
the left-handed fields, and $V_{ij}$ are elements of the CKM matrix
$V$. In general $V_{ij}$ are complex numbers, but their absolute
values are measurable quantities. For example, $|V_{cb}|$ primarily
determines the lifetime of $B$ mesons. The phases of $V_{ij}$,
however, are not physical, like the phases of quark fields. A phase
transformation of the $u$ quark ($u \rightarrow u ~ e^{{\rm
i}\alpha}$), for example, leaves the quark mass term invariant but
changes the elements in the first row of $V$ (i.e., $V_{uj} \rightarrow 
V_{uj} ~ e^{-{\rm i}\alpha}$). Only a common phase transformation of all 
quark fields leaves all elements of $V$ invariant, thus there is a
five-fold freedom to adjust the phases of $V_{ij}$.

In general the unitary matrix $V$ depends on nine parameters.
Note that in the absence of complex phases $V$ would consist of only three 
independent parameters, corresponding to three (Euler) rotation
angles. Hence one can describe the complex matrix $V$ by three
angles and six phases. Due to the freedom in redefining the quark
field phases, five of the six phases in $V$ can be absorbed; and we arrive
at the well-known result that the CKM matrix $V$ can be parametrized
in terms of three rotation angles and one $CP$-violating phase. The
question about how many different ways to describe $V$ may exist was
raised some time ago \cite{Jarlskog89}. Below we shall
reconsider this problem and give a complete analysis.

If the flavor mixing matrix $V$ is first assumed to be a real orthogonal matrix, it can
in general be written as a product of three matrices $R_{12}$,
$R_{23}$ and $R_{31}$, which describe simple rotations in the (1,2),
(2,3) and (3,1) planes:
\begin{eqnarray}
R_{12}(\theta) & = & \left ( \matrix{
c^{~}_{\theta}	& s^{~}_{\theta}	& 0 \cr
- s^{~}_{\theta} 	& c^{~}_{\theta}	& 0 \cr
0	& 0	& 1 \cr} \right ) \; , \nonumber \\ \nonumber \\
R_{23}(\sigma) & = & \left ( \matrix{
1	& 0	& 0 \cr
0 	& c_{\sigma}	& s_{\sigma} \cr
0	& - s_{\sigma}	& c_{\sigma} \cr} \right ) \; , \nonumber \\
\nonumber \\
R_{31}(\tau) & = & \left ( \matrix{
c_{\tau}	& 0	& s_{\tau} \cr
0 	& 1	& 0 \cr
- s_{\tau}	& 0	& c_{\tau} \cr} \right ) \; ,
\end{eqnarray}
where $s^{~}_{\theta} \equiv \sin \theta$, $c^{~}_{\theta} \equiv \cos
\theta$, etc.
Clearly any two rotation matrices do not commute with each other.
There exist twelve different ways to arrange products of these
matrices such that the most general orthogonal matrix $R$ can be
obtained \cite{Jarlskog89}. 
Note that the matrix $R^{-1}_{ij} (\omega) $ plays an equivalent role
as $R_{ij} (\omega) $ in constructing $R$, because of $R^{-1}_{ij}(\omega) =
R_{ij}(-\omega)$. Note also that $R_{ij} (\omega) R_{ij}
(\omega^{\prime}) = R_{ij} (\omega + \omega^{\prime})$ holds, thus 
the product $R_{ij}(\omega) R_{ij}(\omega^{\prime})
R_{kl}(\omega^{\prime\prime})$ or $R_{kl}(\omega^{\prime\prime})
R_{ij}(\omega) R_{ij}(\omega^{\prime})$ cannot cover the whole space
of a $3\times 3$ orthogonal matrix and should be excluded.
Explicitly the twelve different forms of $R$ read as
\begin{eqnarray}
(1) & & R \; =\; R_{12}(\theta) ~ R_{23}(\sigma) ~ R_{12}(\theta^{\prime})
\; , \nonumber \\
(2) & & R \; =\; R_{12}(\theta) ~ R_{31}(\tau) ~ R_{12}(\theta^{\prime})
\; , \nonumber \\
(3) & & R \; =\; R_{23}(\sigma) ~ R_{12}(\theta) ~ R_{23}(\sigma^{\prime})
\; , \nonumber \\
(4) & & R \; =\; R_{23}(\sigma) ~ R_{31}(\tau) ~ R_{23}(\sigma^{\prime})
\; , \nonumber \\
(5) & & R \; =\; R_{31}(\tau) ~ R_{12}(\theta) ~ R_{31}(\tau^{\prime})
\; , \nonumber \\
(6) & & R \; =\; R_{31}(\tau) ~ R_{23}(\sigma) ~ R_{31}(\tau^{\prime})
\; , \nonumber
\end{eqnarray}
in which a rotation in the $(i,j)$ plane occurs twice;
and
\begin{eqnarray}
(7) & & R \; =\; R_{12}(\theta) ~ R_{23}(\sigma) ~ R_{31}(\tau)
\; , \nonumber \\
(8) & & R \; =\; R_{12}(\theta) ~ R_{31}(\tau) ~ R_{23}(\sigma)
\; , \nonumber \\
(9) & & R \; =\; R_{23}(\sigma) ~ R_{12}(\theta) ~ R_{31}(\tau)
\; , \nonumber \\
(10) & & R \; =\; R_{23}(\sigma) ~ R_{31}(\tau) ~ R_{12}(\theta)
\; , \nonumber \\
(11) & & R \; =\; R_{31}(\tau) ~ R_{12}(\theta) ~ R_{23}(\sigma)
\; , \nonumber \\
(12) & & R \; =\; R_{31}(\tau) ~ R_{23}(\sigma) ~ R_{12}(\theta) 
\; , \nonumber 
\end{eqnarray}
where all three $R_{ij}$ are present.

Although all the above twelve combinations represent the most general 
orthogonal matrices, only nine of them are structurally different.
The reason is that the products $R_{ij} R_{kl} R_{ij}$ and $R_{ij} R_{mn} R_{ij}$ (with
$ij\neq kl\neq mn$) are correlated with each other, leading
essentially to the same form for $R$. Indeed it is straightforward to
see the correlation between patterns (1), (3), (5) and (2), (4),
(6), respectively, as follows:
\begin{eqnarray}
R_{12}(\theta) ~ R_{31}(\tau) ~ R_{12}(\theta^{\prime})
& = & R_{12}(\theta + \pi/2) ~ R_{23}(\sigma = \tau) ~
R_{12}(\theta^{\prime} - \pi/2) \; , \nonumber \\
R_{23}(\sigma) ~ R_{31}(\tau) ~ R_{23}(\sigma^{\prime})
& = & R_{23}(\sigma -\pi/2) ~ R_{12}(\theta = \tau) ~
R_{23}(\sigma^{\prime} + \pi/2) \; , \nonumber \\
R_{31}(\tau) ~ R_{23}(\sigma) ~ R_{31}(\tau^{\prime})
& = & R_{31}(\tau  + \pi/2) ~ R_{12}(\theta = \sigma) ~
R_{31}(\tau^{\prime} - \pi/2) \; .
\end{eqnarray}
Thus the orthogonal matrices (2), (4) and (6) need not be treated as
independent choices. 
We then draw the conclusion that
there exist {\it nine} different forms for the orthogonal matrix $R$,
i.e., patterns (1), (3) and (5) as well as (7) -- (12).

We proceed to include the $CP$-violating phase, denoted by $\varphi$,
in the above rotation matrices. The resultant matrices should be
unitary such that a unitary flavor mixing matrix can be finally
produced. There are several different ways for
$\varphi$ to enter $R_{12}$, e.g., 
$$
R_{12} (\theta, \varphi) \; =\; \left ( \matrix{
c^{~}_{\theta}	& s^{~}_{\theta} ~ e^{+{\rm i} \varphi}	& 0 \cr
- s^{~}_{\theta} ~ e^{-{\rm i} \varphi} 	& c^{~}_{\theta}	& 0 \cr
0	& 0	& 1 \cr} \right ) \; , 
\eqno(4{\rm a})
$$
or
$$
R_{12} (\theta, \varphi) \; =\; \left ( \matrix{
c^{~}_{\theta}	& s^{~}_{\theta} 	& 0 \cr
- s^{~}_{\theta}  	& c^{~}_{\theta}	& 0 \cr
0	& 0	& e^{-{\rm i} \varphi} \cr} \right ) \; , 
\eqno(4{\rm b})
$$
or
$$
R_{12} (\theta, \varphi) \; =\; \left ( \matrix{
c^{~}_{\theta} ~ e^{+{\rm i} \varphi}	& s^{~}_{\theta} 	& 0 \cr
- s^{~}_{\theta}  	& c^{~}_{\theta} ~ e^{-{\rm i} \varphi}	& 0 \cr
0	& 0	& 1 \cr} \right ) \; .
\eqno(4{\rm c})
$$
Similarly one may introduce a phase parameter into $R_{23}$ or
$R_{31}$. Then the CKM matrix $V$ can be constructed, as a product of
three rotation matrices, by use of one complex $R_{ij}$ and two real ones. 
Note that the location of the $CP$-violating phase in $V$ can be arranged by
redefining the quark field phases, thus it does not play an essential role in 
classifying different parametrizations. We find that it is always
possible to locate the phase parameter $\varphi$ in a $2\times 2$ submatrix of
$V$, in which each element is a sum of two terms with the relative
phase $\varphi$. The remaining five elements of $V$ are real in such a 
phase assignment. Accordingly we arrive at nine distinctive
parametrizations of the CKM matrix $V$, as listed in Table 1, where
the complex rotation matrices $R_{12}(\theta, \varphi)$,
$R_{23}(\sigma, \varphi)$ and $R_{31}(\tau, \varphi)$ are obtained
directly from the real ones in Eq. (2) with the replacement $1
\rightarrow e^{-{\rm i}\varphi}$.

Some instructive relations of each parametrization, as well as the
rephasing-invariant measure of $CP$ violation \cite{Jarlskog85} defined by $\cal J$
through
\setcounter{equation}{4}
\begin{equation}
{\rm Im} \left ( V_{il} V_{jm} V^*_{im} V^*_{jl} \right ) \; =\;
{\cal J} \sum_{k,n=1}^{3} \left ( \epsilon^{~}_{ijk} \epsilon^{~}_{lmn} \right ) \; ,
\end{equation}
have also been given in Table 1. One can see that {\it P2} and {\it P3}
correspond to the Kobayashi-Maskawa \cite{KM73} and Maiani
\cite{Others} representations, although different
notations for the $CP$-violating phase and three mixing angles are adopted
here. The latter is indeed equivalent to the 
``standard'' parametrization advocated by the Particle Data Group
\cite{Others,PDG96}. This can be seen clearly if one makes 
three transformations of quark field phases: 
$c \rightarrow c ~ e^{-{\rm i} \varphi}$, $t
\rightarrow t ~ e^{-{\rm i} \varphi}$, and $b \rightarrow 
b ~ e^{-{\rm i} \varphi}$. In addition, {\it P1} is just the one proposed by the
present authors in Ref. \cite{FX97}.

{\rm From} a mathematical point of view, all nine different parametrizations
are equivalent. However this is not the case if we apply our
considerations to the quarks and their mass spectrum. It is well known that both the observed
quark mass spectrum and the observed values of the flavor mixing
parameters exhibit a striking hierarchical structure. The latter can
be understood in a natural way as the consequence of a specific
pattern of chiral symmetries whose breaking causes the towers of
different masses to appear step by step
\cite{Fritzsch87a,Fritzsch87b,Hall93}. Such a chiral evolution of the
mass matrices leads, as argued in Ref. \cite{Fritzsch87b}, to a
specific way to introduce and describe the flavor mixing. In the limit 
$m_u = m_d =0$, which is close to the real world, since $m_u/m_t \ll 1$ and
$m_d/m_b \ll 1$, the flavor mixing is merely a rotation between the
$t$--$c$ and $b$--$s$ systems, described by one rotation angle. No complex
phase is present; i.e., $CP$ violation is absent. This rotation angle
is expected to change very little, once $m_u$ and $m_d$
are introduced as tiny perturbations. A sensible parametrization should
make use of this feature. This implies that the rotation matrix
$R_{23}$ appears exactly once in the description of the CKM matrix
$V$, eliminating {\it P2} (in which $R_{23}$ appears twice) and {\it P5}
(where $R_{23}$ is absent). This leaves us with seven parametrizations 
of the flavor mixing matrix.

The list can be reduced further by considering the location of the phase $\varphi$. In
the limit $m_u = m_d =0$, the phase must disappear in the weak
transition elements $V_{tb}$, $V_{ts}$, $V_{cb}$ and $V_{cs}$. In
{\it P7} and {\it P8}, however, $\varphi$ appears particularly in
$V_{tb}$. Thus these two parametrizations should be eliminated, leaving
us with five parametrizations (i.e., {\it P1}, {\it P3}, {\it P4}, {\it P6} and
{\it P9}). In the same limit, the phase $\varphi$ appears in the $V_{ts}$
element of {\it P3} and the $V_{cb}$ element of {\it P4}. Hence these two
parametrizations should also be eliminated. Then we are left with three
parametrizations, {\it P1}, {\it P6} and {\it P9}. As expected, these are the
parametrizations containing the complex rotation matrix
$R_{23}(\sigma, \varphi)$. We stress that the ``standard'' parametrization
\cite{PDG96} (equivalent to {\it P3}) does not obey the above constraints and
should be dismissed.

Among the remaining three parametrizations, {\it P1} is singled out by 
the fact that the $CP$-violating phase $\varphi$ appears only in the
$2\times 2$ submatrix of $V$ describing the weak transitions among the 
light quarks. This is precisely the system where the phase $\varphi$
should appear, not in any of the weak transition elements involving the 
heavy quarks $t$ and $b$.

In the parametrization {\it P6} or {\it P9}, the complex
phase $\varphi$ appears in $V_{cb}$ or $V_{ts}$, but this phase factor
is multiplied by a product of $\sin\theta$ and $\sin\tau$, i.e., it is of second 
order of the weak mixing angles. Hence the imaginary parts of these
elements are not exactly vanishing, but very small in magnitude.

In our view the best possibility to describe the flavor mixing in the
standard model is to adopt the parametrization {\it P1}. As discussed
in Ref. \cite{FX97}, this parametrization has a number of significant
advantages in addition to that mentioned above. Especially it is well
suited for specific models of quark mass matrices (see, e.g., Refs. 
\cite{Fritzsch79,Hall}).

We conclude: there are nine different ways to describe a real $3\times
3$ flavor mixing matrix in terms of three rotation angles. Introducing 
a complex phase $\varphi$ does not increase the number of distinct
parametrizations, except for the fact that there is a continuum of
possibilities for assigning the phase factors. Imposing natural
constraints in view of the observed mass hierarchy (i.e., 
in the limit $m_u = m_d =0$ phases should be absent in the (2,2), (2,3), (3,2) and (3,3)
elements of the mixing matrix), we can eliminate six parametrizations, including the
original Kobayashi-Maskawa parametrization \cite{KM73} and the
``standard'' parametrization proposed in Refs. \cite{Others,PDG96}. We propose to use the
parametrization {\it P1} for the
further study of flavor mixing and $CP$-violating phenomena.

{\it Acknowledgments:} ~ The work of H.F. was supported in part by 
GIF contract I-0304-120.07/93 and
EEC contract CHRX-CT94-0579 (DG 12 COMA). Z.Z.X. is grateful to
A.I. Sanda for his warm hospitality and to the Japan Society for the
Promotion of Science for its financial support.

{\it Note added:} ~ After completion of this work we received a
preprint of A. Rasin \cite{Rasin97}, in which part of the conclusions
drawn here was also reached.

\newpage

\renewcommand{\baselinestretch}{1.09}
\small 
\begin{table}
\caption{Classification of different parametrizations for the flavor mixing
matrix.}
\vspace{-0.5cm}
\begin{center}
\begin{tabular}{ccc} \\ \hline\hline 
Parametrization     & ~~~~ & Useful relations 
\\  \hline \\
{\it P1:} ~ $V \; = \; R_{12}(\theta) ~ R_{23}(\sigma, \varphi)
~ R^{-1}_{12}(\theta^{\prime})$ 	
&& ${\cal J} = s^{~}_{\theta} c^{~}_{\theta} s^{~}_{\theta^{\prime}}
c^{~}_{\theta^{\prime}} s^2_{\sigma} c_{\sigma} \sin\varphi$ \\
$\left ( \matrix{
s^{~}_{\theta} s^{~}_{\theta^{\prime}} c_{\sigma} + c^{~}_{\theta} c^{~}_{\theta^{\prime}} e^{-{\rm i}\varphi} 	& 
s^{~}_{\theta} c^{~}_{\theta^{\prime}} c_{\sigma} - c^{~}_{\theta}
s^{~}_{\theta^{\prime}} e^{-{\rm i}\varphi}	& s^{~}_{\theta} s_{\sigma} \cr
c^{~}_{\theta} s^{~}_{\theta^{\prime}} c_{\sigma} - s^{~}_{\theta} c^{~}_{\theta^{\prime}} e^{-{\rm i}\varphi}	&
c^{~}_{\theta} c^{~}_{\theta^{\prime}} c_{\sigma} + s^{~}_{\theta}
s^{~}_{\theta^{\prime}} e^{-{\rm i}\varphi}	& c^{~}_{\theta} s_{\sigma} \cr
- s^{~}_{\theta^{\prime}} s_{\sigma} 	& - c^{~}_{\theta^{\prime}}
s_{\sigma}	& c_{\sigma} \cr} \right )
$
&& $\matrix{
\tan\theta = |V_{ub}/V_{cb}| \cr
\tan\theta^{\prime} = |V_{td}/V_{ts}| \cr
\cos\sigma = |V_{tb}| \cr} $ \\ \\
{\it P2:} ~ $V \; = \; R_{23}(\sigma) ~ R_{12}(\theta, \varphi)
~ R^{-1}_{23}(\sigma^{\prime})$ 	
&& ${\cal J} = s^2_{\theta} c^{~}_{\theta} s_{\sigma} c_{\sigma} s_{\sigma^{\prime}} c_{\sigma^{\prime}} \sin\varphi$ \\ 
$\left ( \matrix{
c^{~}_{\theta} 	& s^{~}_{\theta} c_{\sigma^{\prime}} 	& -s^{~}_{\theta} s_{\sigma^{\prime}} \cr 
-s^{~}_{\theta} c_{\sigma} 	& c^{~}_{\theta} c_{\sigma} c_{\sigma^{\prime}} + s_{\sigma} s_{\sigma^{\prime}} e^{-{\rm i}\varphi}	
& -c^{~}_{\theta} c_{\sigma} s_{\sigma^{\prime}} + s_{\sigma} c_{\sigma^{\prime}} e^{-{\rm i}\varphi} \cr
s^{~}_{\theta} s_{\sigma} 	& -c^{~}_{\theta} s_{\sigma} c_{\sigma^{\prime}} + c_{\sigma} s_{\sigma^{\prime}} e^{-{\rm i}\varphi}	
& c^{~}_{\theta} s_{\sigma} s_{\sigma^{\prime}} + c_{\sigma} c_{\sigma^{\prime}} e^{-{\rm i}\varphi} \cr} \right )
$
&& $\matrix{
\cos\theta = |V_{ud}| \cr
\tan\sigma = |V_{td}/V_{cd}| \cr
\tan\sigma^{\prime} = |V_{ub}/V_{us}| \cr} $ \\ \\
{\it P3:} ~ $V \; = \; R_{23}(\sigma) ~ R_{31}(\tau, \varphi)
~ R_{12}(\theta)$ 	
&& ${\cal J} = s^{~}_{\theta} c^{~}_{\theta} s_{\sigma} c_{\sigma} s_{\tau} c^2_{\tau} \sin\varphi$ 
\\ 
$\left ( \matrix{
c^{~}_{\theta} c_{\tau} 	& s^{~}_{\theta} c_{\tau} 	& s_{\tau} \cr 
-c^{~}_{\theta} s_{\sigma} s_{\tau} - s^{~}_{\theta} c_{\sigma} e^{-{\rm i}\varphi}	
& -s^{~}_{\theta} s_{\sigma} s_{\tau} + c^{~}_{\theta} c_{\sigma} e^{-{\rm i}\varphi} 	& s_{\sigma} c_{\tau} \cr
-c^{~}_{\theta} c_{\sigma} s_{\tau} + s^{~}_{\theta} s_{\sigma} e^{-{\rm i}\varphi}	
& -s^{~}_{\theta} c_{\sigma} s_{\tau} - c^{~}_{\theta} s_{\sigma} e^{-{\rm i}\varphi} 	& c_{\sigma} c_{\tau} \cr} \right )
$
&& $\matrix{
\tan\theta = |V_{us}/V_{ud}| \cr
\tan\sigma = |V_{cb}/V_{tb}| \cr
\sin\tau = |V_{ub}| \cr} $ \\ \\
{\it P4:} ~ $V \; = \; R_{12}(\theta) ~ R_{31}(\tau, \varphi)
~ R^{-1}_{23}(\sigma)$ 	
&& ${\cal J} = s^{~}_{\theta} c^{~}_{\theta} s_{\sigma} c_{\sigma} s_{\tau} c^2_{\tau} \sin\varphi$ 
\\
$\left ( \matrix{
c^{~}_{\theta} c_{\tau} 	& c^{~}_{\theta} s_{\sigma} s_{\tau} + s^{~}_{\theta} c_{\sigma} e^{-{\rm i}\varphi}	
& c^{~}_{\theta} c_{\sigma} s_{\tau} - s^{~}_{\theta} s_{\sigma} e^{-{\rm i}\varphi} \cr
-s^{~}_{\theta} c_{\tau} 	& -s^{~}_{\theta} s_{\sigma} s_{\tau} + c^{~}_{\theta} c_{\sigma} e^{-{\rm i}\varphi}	
& -s^{~}_{\theta} c_{\sigma} s_{\tau} - c^{~}_{\theta} s_{\sigma} e^{-{\rm i}\varphi} \cr
-s_{\tau}	& s_{\sigma} c_{\tau}	& c_{\sigma} c_{\tau} \cr} \right )
$
&& $\matrix{
\tan\theta = |V_{cd}/V_{ud}| \cr
\tan\sigma = |V_{ts}/V_{tb}| \cr
\sin\tau = |V_{td}| \cr} $ \\ \\
{\it P5:} ~ $V \; = \; R_{31}(\tau) ~ R_{12}(\theta, \varphi)
~ R^{-1}_{31}(\tau^{\prime})$ 	
&& ${\cal J} = s^2_{\theta} c^{~}_{\theta} s_{\tau} c_{\tau} s_{\tau^{\prime}} c_{\tau^{\prime}} \sin\varphi$ 
\\
$\left ( \matrix{
c^{~}_{\theta} c_{\tau} c_{\tau^{\prime}} + s_{\tau} s_{\tau^{\prime}} e^{-{\rm i}\varphi}	& s^{~}_{\theta} c_{\tau}
& -c^{~}_{\theta} c_{\tau} s_{\tau^{\prime}} + s_{\tau} c_{\tau^{\prime}} e^{-{\rm i}\varphi} \cr
-s^{~}_{\theta} c_{\tau^{\prime}} 	& c^{~}_{\theta}	& s^{~}_{\theta} s_{\tau^{\prime}} \cr 
-c^{~}_{\theta} s_{\tau} c_{\tau^{\prime}} + c_{\tau} s_{\tau^{\prime}} e^{-{\rm i}\varphi}	& -s^{~}_{\theta} s_{\tau}
& c^{~}_{\theta} s_{\tau} s_{\tau^{\prime}} + c_{\tau} c_{\tau^{\prime}} e^{-{\rm i}\varphi} \cr} \right )
$
&& $\matrix{
\cos\theta = |V_{cs}| \cr
\tan\tau = |V_{ts}/V_{us}| \cr
\tan\tau^{\prime} = |V_{cb}/V_{cd}| \cr} $ \\ \\
{\it P6:} ~ $V \; = \; R_{12}(\theta) ~ R_{23}(\sigma, \varphi)
~ R_{31}(\tau)$ 	
&& ${\cal J} = s^{~}_{\theta} c^{~}_{\theta} s_{\sigma} c^2_{\sigma} s_{\tau} c_{\tau} \sin\varphi$ 
\\
$\left ( \matrix{
-s^{~}_{\theta} s_{\sigma} s_{\tau} + c^{~}_{\theta} c_{\tau} e^{-{\rm i}\varphi} 	& s^{~}_{\theta} c_{\sigma} 	&
s^{~}_{\theta} s_{\sigma} c_{\tau} + c^{~}_{\theta} s_{\tau} e^{-{\rm i}\varphi} \cr
-c^{~}_{\theta} s_{\sigma} s_{\tau} - s^{~}_{\theta} c_{\tau} e^{-{\rm i}\varphi}	& c^{~}_{\theta} c_{\sigma}	&
c^{~}_{\theta} s_{\sigma} c_{\tau} - s^{~}_{\theta} s_{\tau} e^{-{\rm i}\varphi} \cr
-c_{\sigma} s_{\tau} 	& -s_{\sigma} 		& c_{\sigma} c_{\tau} \cr} \right )
$
&& $\matrix{
\tan\theta = |V_{us}/V_{cs}| \cr
\sin\sigma = |V_{ts}| \cr
\tan\tau = |V_{td}/V_{tb}| \cr} $ \\ \\
{\it P7:} ~ $V \; = \; R_{23}(\sigma) ~ R_{12}(\theta, \varphi)
~ R^{-1}_{31}(\tau)$ 	
&& ${\cal J} = s^{~}_{\theta} c^2_{\theta} s_{\sigma} c_{\sigma} s_{\tau} c_{\tau} \sin\varphi$ 
\\
$\left ( \matrix{
c^{~}_{\theta} c_{\tau} 	& s^{~}_{\theta}  	& -c^{~}_{\theta} s_{\tau} \cr 
-s^{~}_{\theta} c_{\sigma} c_{\tau} + s_{\sigma} s_{\tau} e^{-{\rm i}\varphi}	& c^{~}_{\theta} c_{\sigma}
& s^{~}_{\theta} c_{\sigma} s_{\tau} + s_{\sigma} c_{\tau} e^{-{\rm i}\varphi} \cr
s^{~}_{\theta} s_{\sigma} c_{\tau} + c_{\sigma} s_{\tau} e^{-{\rm i}\varphi}	& -c^{~}_{\theta} s_{\sigma}	
& -s^{~}_{\theta} s_{\sigma} s_{\tau} + c_{\sigma} c_{\tau} e^{-{\rm i}\varphi} \cr} \right )
$
&& $\matrix{
\sin\theta = |V_{us}| \cr
\tan\sigma = |V_{ts}/V_{cs}| \cr
\tan\tau = |V_{ub}/V_{ud}| \cr} $ \\ \\
{\it P8:} ~ $V \; = \; R_{31}(\tau) ~ R_{12}(\theta, \varphi)
~ R_{23}(\sigma)$ 	
&& ${\cal J} = s^{~}_{\theta} c^2_{\theta} s_{\sigma} c_{\sigma} s_{\tau} c_{\tau} \sin\varphi$ 
\\
$\left ( \matrix{
c^{~}_{\theta} c_{\tau} 	& s^{~}_{\theta} c_{\sigma} c_{\tau} - s_{\sigma} s_{\tau} e^{-{\rm i}\varphi}	
& s^{~}_{\theta} s_{\sigma} c_{\tau} + c_{\sigma} s_{\tau} e^{-{\rm i}\varphi} \cr
-s^{~}_{\theta}	& c^{~}_{\theta} c_{\sigma}	& c^{~}_{\theta} s_{\sigma} \cr
-c^{~}_{\theta} s_{\tau}	& -s^{~}_{\theta} c_{\sigma} s_{\tau} - s_{\sigma} c_{\tau} e^{-{\rm i}\varphi}	
& -s^{~}_{\theta} s_{\sigma} s_{\tau} + c_{\sigma} c_{\tau} e^{-{\rm i}\varphi} \cr} \right )
$
&& $\matrix{
\sin\theta = |V_{cd}| \cr
\tan\sigma = |V_{cb}/V_{cs}| \cr
\tan\tau = |V_{td}/V_{ud}| \cr} $ \\ \\
{\it P9:} ~ $V \; = \; R_{31}(\tau) ~ R_{23}(\sigma, \varphi)
~ R^{-1}_{12}(\theta)$ 	
&& ${\cal J} = s^{~}_{\theta} c^{~}_{\theta} s_{\sigma} c^2_{\sigma} s_{\tau} c_{\tau} \sin\varphi$ 
\\
$\left ( \matrix{
-s^{~}_{\theta} s_{\sigma} s_{\tau} + c^{~}_{\theta} c_{\tau} e^{-{\rm i}\varphi}	
& -c^{~}_{\theta} s_{\sigma} s_{\tau} - s^{~}_{\theta} c_{\tau} e^{-{\rm i}\varphi} 	& c_{\sigma} s_{\tau} \cr
s^{~}_{\theta} c_{\sigma} 	& c^{~}_{\theta} c_{\sigma}	& s_{\sigma} \cr 
-s^{~}_{\theta} s_{\sigma} c_{\tau} - c^{~}_{\theta} s_{\tau} e^{-{\rm i}\varphi}	
& -c^{~}_{\theta} s_{\sigma} c_{\tau} + s^{~}_{\theta} s_{\tau} e^{-{\rm i}\varphi} 	& c_{\sigma} c_{\tau} \cr} \right )
$
&& $\matrix{
\tan\theta = |V_{cd}/V_{cs}| \cr
\sin\sigma = |V_{cb}| \cr
\tan\tau = |V_{ub}/V_{tb}| \cr} $ \\ \\
\hline\hline
\end{tabular}
\end{center}
\end{table}

\end{document}